\documentclass[a4paper,10pt]{extarticle}
\pdfoutput=1 

\usepackage{jheppub} 
\usepackage{hyperref}
\hypersetup{
    bookmarks=true,         
    unicode=false,          
    pdftoolbar=true,        
   pdfmenubar=true,        
    pdffitwindow=false,     
    pdfstartview={FitH},    
    pdftitle={My title},    
    pdfauthor={Author},     
    pdfsubject={Subject},   
    pdfcreator={Creator},   
    pdfproducer={Producer}, 
    pdfkeywords={keyword1} {key2} {key3}, 
    pdfnewwindow=true,      
    colorlinks=true,       
    linkcolor=red,          
    citecolor=purple,        
    filecolor=magenta,      
    urlcolor=blue,           
    linktocpage=true
}
\usepackage{graphics}
\usepackage{rotating}
\usepackage{subfigure}
\usepackage{pstricks}
\usepackage{color}
\usepackage{amsfonts}
\usepackage{mathrsfs,amsmath}
\usepackage{epsfig}
\usepackage{amsmath,amssymb,amsthm,graphicx,latexsym}
\usepackage{ulem}

\newcommand{\be}{\begin{equation}}
\newcommand{\ee}{\end{equation}}
\newcommand{\bea}{\begin{eqnarray}}
\newcommand{\eea}{\end{eqnarray}}
\newcommand{\bml}{\begin{subequations}}
\newcommand{\eml}{\end{subequations}}
\newcommand{\bfig}{\begin{figure}}
\newcommand{\efig}{\end{figure}}

\newcommand{\slashed}{\hspace{-1.1ex}/}

\newcommand{\bmat}{\begin{pmatrix}}
\newcommand{\emat}{\end{pmatrix}}
\begin{document}
	$~~~~~~~~~~~~~~~~~~~~~~~~~~~~~~~~~~~~~~~~~~~~~~~~~~~~~~~~~~~~~~~~~~~~~~~~~~~~~~~~~~~~~~~~~~~~~~~~~~~~~~~~~~~~~~~~~~~~~~~$\textcolor{red}{\bf \Large TIFR/TH/16-49}
	\title{\textsc{\fontsize{47}{90}\selectfont \sffamily \bfseries
	\textcolor{purple}{Bell violation in primordial cosmology}}}

	\author[a]{Sayantan Choudhury,
		\footnote{\textcolor{purple}{\bf Presently working as a Visiting (Post-Doctoral) fellow at DTP, TIFR, Mumbai, Alternative
	E-mail: sayanphysicsisi@gmail.com}. This article is based on talk presented at Varying Constants and Fundamental Cosmology-VARCOSMOFUN'16.${}^{}$} }
	\author[b,c]{Sudhakar Panda
}
	\author[d]{and Rajeev Singh}
	
	\affiliation[a]{Department of Theoretical Physics, Tata Institute of Fundamental Research, Colaba, Mumbai - 400005, India.
	}
	\affiliation[b]{Institute of Physics, Sachivalaya Marg, Bhubaneswar, Odisha - 751005, India.
	}
	\affiliation[c]{Homi Bhabha National Institute, Training School Complex,
	Anushakti Nagar, Mumbai-400085, India.
	}
	\affiliation[d]{Department of Physics, Savitribai Phule Pune University, Pune - 411007, India.
	}
	\emailAdd{sayantan@theory.tifr.res.in, panda@iopb.res.in, rajeevsingh240291@gmail.com }

	\abstract{In this paper, we have worked on the possibility of setting up an Bell's inequality violating experiment
in the context of primordial cosmology following the fundamental principles of quantum mechanics.
To set up this proposal we have introduced a model independent
theoretical framework using which we have studied
the creation of new massive particles for the scalar 
fluctuations in the presence of additional time dependent mass parameter.
Next we explicitly computed the one point and two point correlation functions from this setup.
Then we comment on the measurement techniques of isospin breaking interactions of newly introduced massive particles and its 
further prospects. After that, we give an example of string theory originated axion monodromy model in this context. Finally, we provide a bound on the heavy particle mass parameter for any arbitrary spin field.}


	\maketitle
	\flushbottom

\section{Introduction}
According to inflationary model, the primordial fluctuations
which we see in CMB (Cosmic Microwave Background) were
produced by quantum mechanical effects in the early universe.
These fluctuations are the origin for the formation of large scale
structure, but the fluctuations we observe at present are actually
classical in nature.
Highly entangled quantum mechanical wave function of the
universe plays a very important role during quantum mechanical
interpretation of the required fluctuations.
Hence, one can use Hartle Hawking Wave function in de-Sitter space.
Thus, due to this fact, quantum mechanical fluctuations can be
theoretically demonstrated and can also be implemented
in the context of primordial cosmology, if and only if we
perform a cosmological experiment using highly entangled quantum mechanical
wave function of the universe which is defined in inflationary period
and eventually violate Bell's inequality.
To describe the background methodology, it is
important to mention that in the context of
quantum mechanics, Bell test experiment can be
described by the measurement of two non-
commutating physical operators. These operators are associated
with two distinct locations in space-time.
Thus, using same analogy in the context of primordial cosmology,
we can perform cosmological observations on two spatially separated
and causally disconnected places upto the epoch of reheating (after inflation).
During these observations we can measure the numerical values of
various cosmological observables (along with cosmic variance),
and can also be computed from scalar curvature fluctuation.
But it is important to note that for all such observations
we cannot measure the value of associated
canonically conjugate momentum.
Hence, for these cosmological observables we cannot
measure the imprints of two non-commuting
operators in primordial cosmology.
But there is subtle point, which is, if these observables
satisfies the minimum requirements of decoherence effect, then possibly
we can perform measurements from two exactly commuting cosmological
observables and therefore we will be able to design a Bell's inequality violating
cosmological experimental setup. We know that in quantum theory, to design such an experimental
setup one has to perform number of repeated measurements
on the same object (which in this context is the same quantum
state of the universe) and therefore in such a physical
situation we can justify the appearance of each
and every measurement using a single quantum
state. In the case of primordial cosmology, we can do the same thing, that is,
consider two spatially separated portions in the sky which plays the same role
of performing repeated cosmological Bell's inequality
violating experiment using the same quantum mechanical state. Therefore
we have the advantage of choosing the required properties of
two spatially separated portions in the sky in order to setup Bell's inequality
violating experimental setup. It's completely possible to setup a Bell's inequality violating
cosmological experimental setup if we can find a link which connect these
non-commutating cosmological observables and classical probability distribution function
originated from model of inflation. In this article, we explore this possibility in detail.

\section{Bell's inequality and its violation in Quantum Mechanics}
{\bf John Stewart Bell} in ref.~\cite{Bell:1964kc} showed that {\it ``In a theory in which parameters are added to quantum mechanics to determine the results of
individual measurements, without changing the statistical predictions, there must be a mechanism whereby the setting of one measuring device can influence the reading of another instrument, however remote.''}. In that paper, J.S Bell gave two assumptions for his theory which were, 1.  concept of reality and 2. concept of locality. Using these assumptions he derived an inequality, which is known as  {\it`Bell's Inequality'}.
In this paper, he proved that {\it ``no local hidden variable theory is compatible with quantum
mechanics.''}
\begin{figure*}[htb]
\centering
{
    \includegraphics[width=8.2cm,height=5cm] {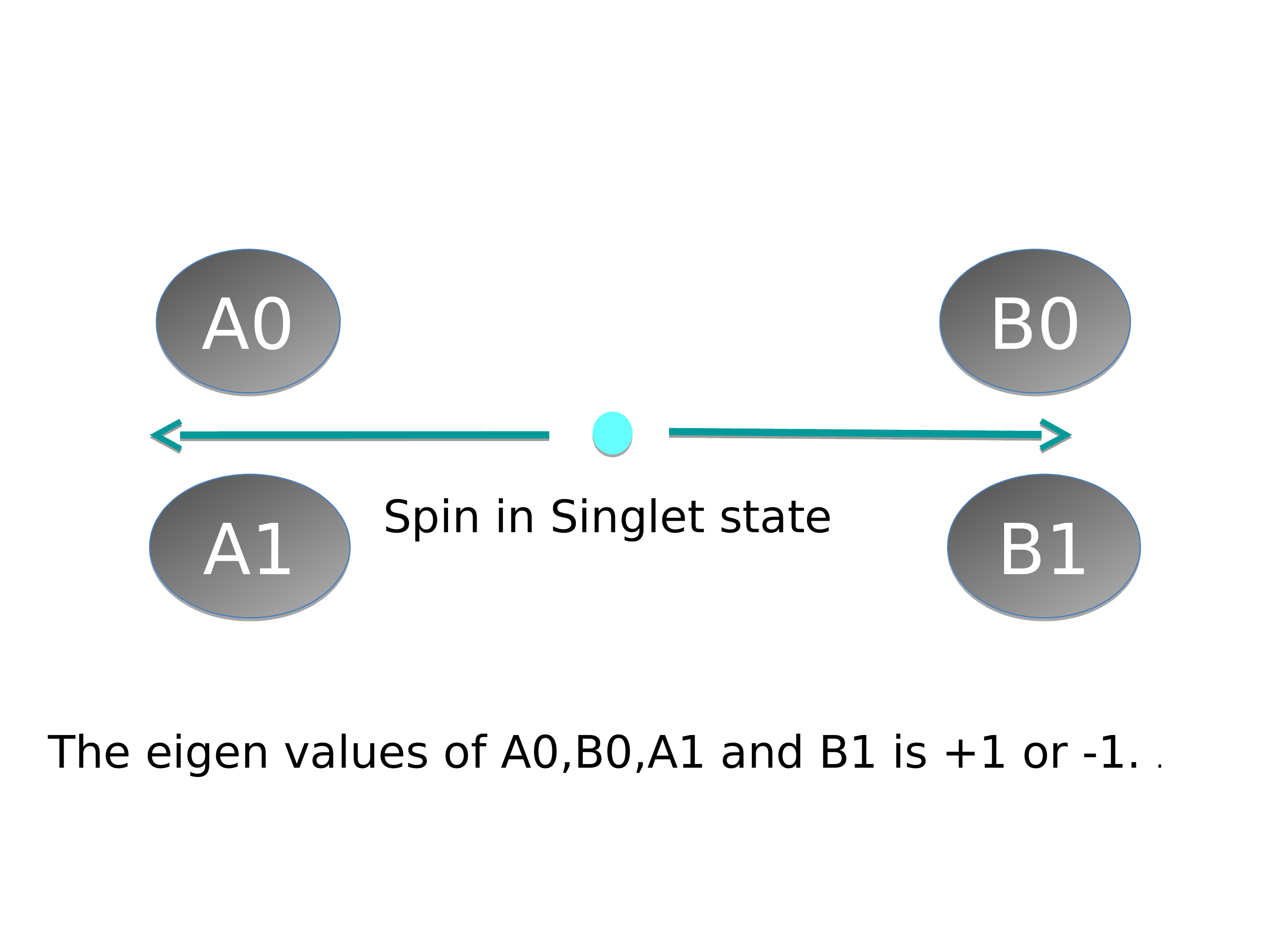}
}
\caption{Schematic diagram of Bell's inequality example for a spin system.}
\label{spinc2xxz}
\end{figure*}
See fig~(\ref{spinc2xxz}) for the representative setup for the spin system.
Here the operators which are $A_0$, $B_0$, $A_1$ and $B_1$ corresponds to measuring the spin and their eigenvalues are $\pm$ 1,
choosing the value of operators as:
\begin{equation} A_0 = {n}_{0}.{\sigma},~~
A_1 = {n}_{1}.{\sigma},~~
B_0 = {n}_{0}.{\sigma},~~
B_1 = {n}_{1}.{\sigma}.\end{equation}
Further considering a variable which is given by:
\begin{equation}\label{qwexxxxx}
\langle R \rangle = \langle A_0 B_0 \rangle + \langle A_1 B_0 \rangle + \langle A_0 B_1 \rangle - \langle A_1 B_1 \rangle
\end{equation}
According to hidden variable theory, $\vert \langle R \rangle \rvert \leq 2$.
But in quantum mechanics, the expectation value of R can be found bigger.
By squaring the Eq.~(\ref{qwexxxxx}) we can show that $R^2 = 4 + [A_1,A_0][B_1,B_0] \Rightarrow \lvert \langle R \rangle \rvert >2$
making $\lvert \langle R \rangle \rvert$ larger than $2$, which violates Bell's inequality. The question now
is, how to draw above conclusion, therefore choosing:
\be A_0 = {x}.{\sigma},~~
A_1 = {y}.{\sigma},~~
B_0 = sin~ \theta ({x}.{\sigma}) + cos~ \theta ({y}.{\sigma}),~~
B_1 = cos~ \theta({x}.{\sigma}) - sin~ \theta ({y}.{\sigma}),\ee
we get the extra $\sqrt{2}$ factor for the maximal violation i.e.
$\lvert \langle R \rangle \rvert >2\sqrt{2}$.
\section{Cosmological Bell violating setup}
\begin{figure*}[htb]
\centering
\subfigure[Successful setup.]{
    \includegraphics[width=7.7cm,height=5cm] {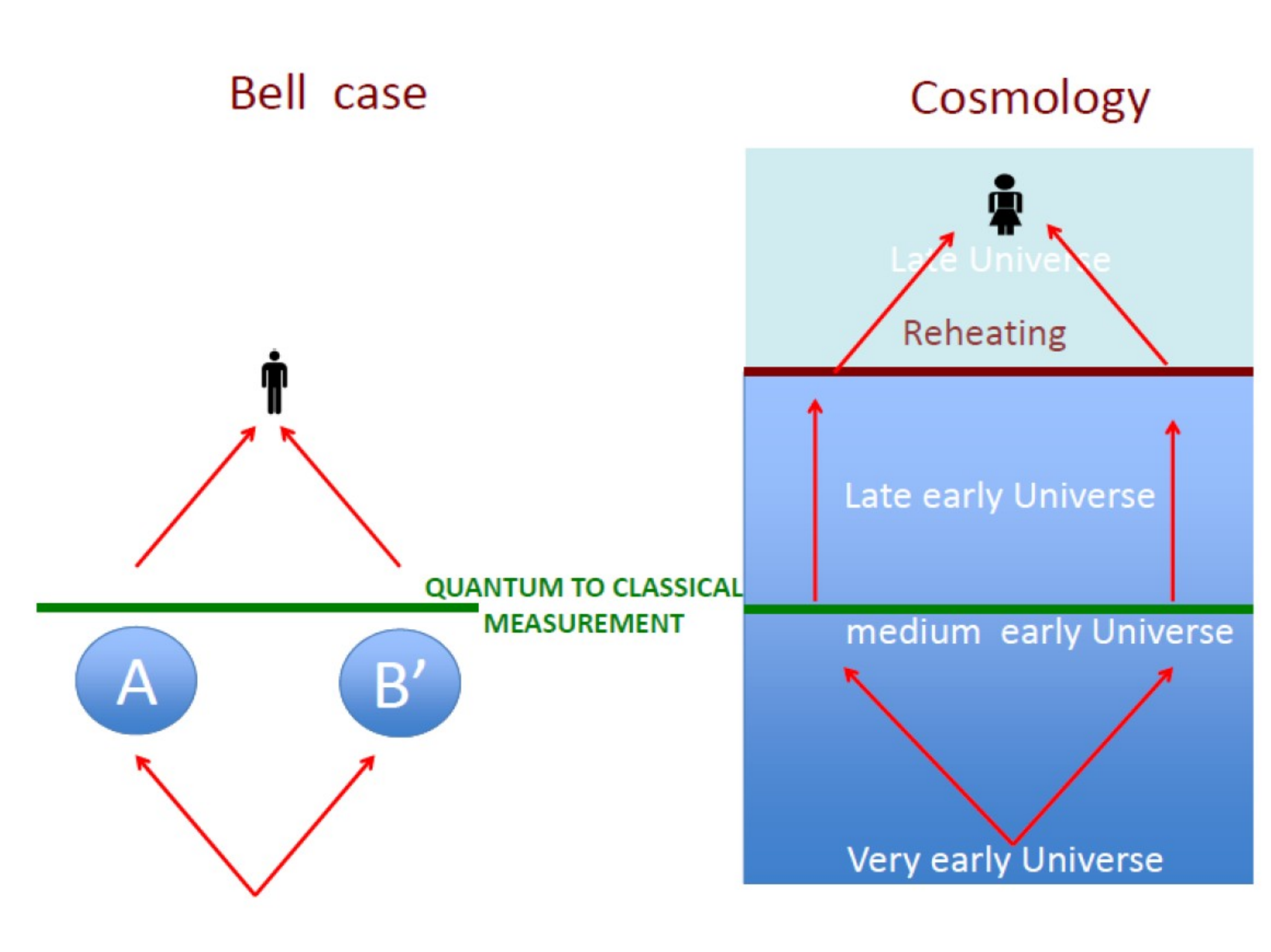}
    \label{fig1xxa}
}
\subfigure[Unsuccessful setup.]{
    \includegraphics[width=7.7cm,height=5cm] {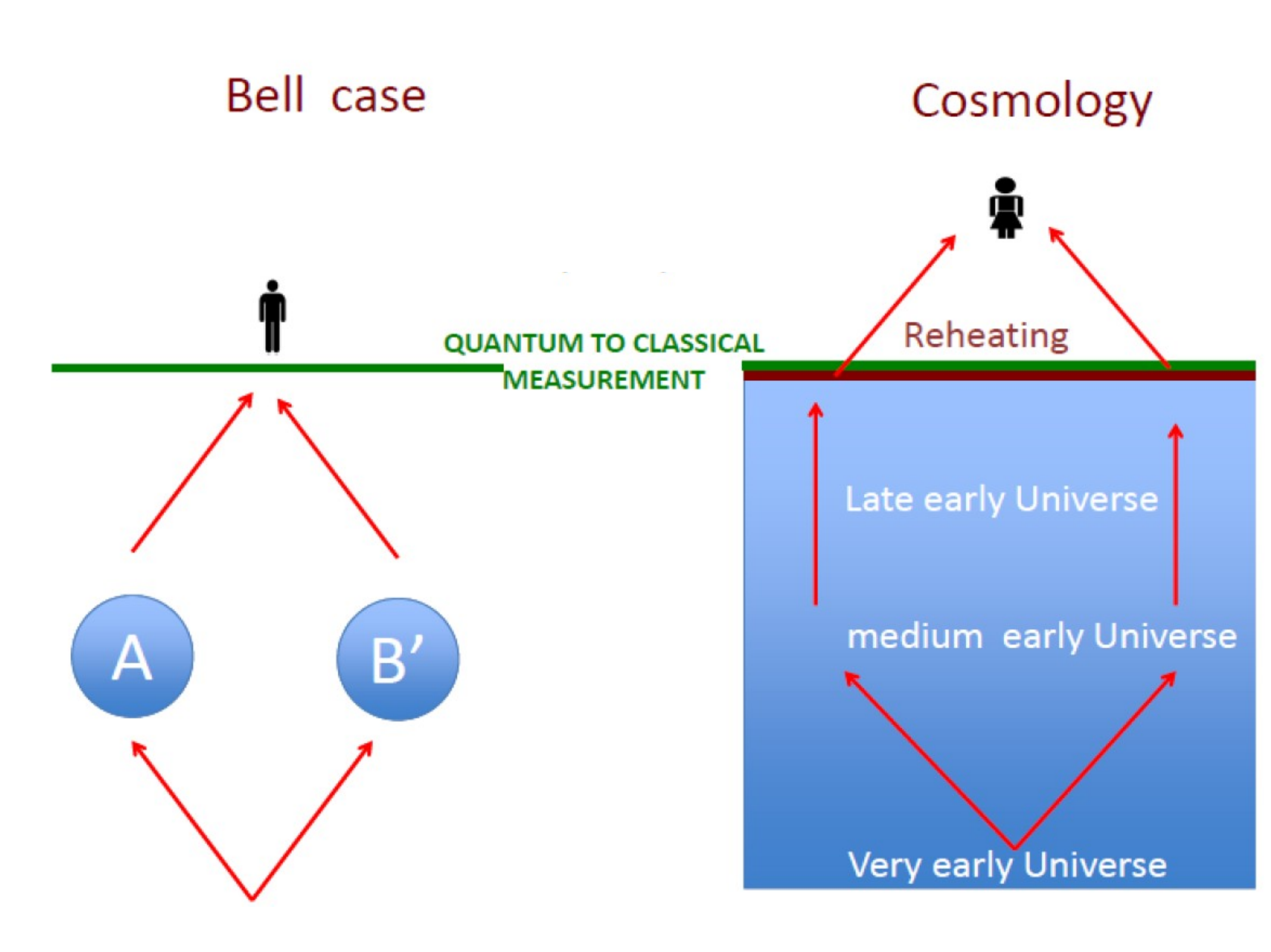}
    \label{fig2xxb}
}
\caption[Optional caption for list of figures]{Successful vs Unsuccessful setup for Bell's inequality violation in Quantum Mechanics and Cosmology \cite{andi}.} 
\label{fzaxx}
\end{figure*}
To describe the setup of Bell violating experiment in the context of cosmology,
we have to start with the quantum fluctuations which can be treated as a time dependent harmonic oscillator in Fourier space which is given by:
\be S=\int \frac{d\eta}{\eta^2}\left(|\partial_{\eta}\phi_{\bf k}(\eta)|^2-k^2|\phi_{\bf k}(\eta)|^2\right).\ee Therefore to describe such quantum fluctuations in expanding space-time we have to use the quasi de Sitter metric, $ds^2=a^2(\eta)(-d\eta^2+d{\bf x}^2),$
where $H$ is the Hubble parameter, ${\bf x}$ represents comoving spatial coordinates, $\eta$ signifies the conformal time coordinate and $a(\eta)=-(1+\epsilon)/H\eta$ is the scale factor with, $\epsilon=-\dot{H}/H^2$. Noting that, here it can easily be shown that quantum theory plays an important role to
produce spatially dependent fluctuations in terms of the scalar fields. See  fig.~(\ref{fzaxx}), where we have shown a comparison between a sucessful and an unsucessful setup for Bell's inequality violation in quantum mechanics and cosmology.

We know that in the early universe, quantum mechanical effects are the seeds
for primordial fluctuations.
But the fluctuations we have observed today is completely classical in nature as we said earlier.
We know that in the theory of inflation 
all such fluctuations become classical in nature as they exit the horizon
and inside the horizon all of them are quantum
in nature. In de Sitter space,
$k^3\left|\left[\phi_{\bf k}(\eta),\eta\partial_{\eta}\phi_{\bf k}\right]\right|\approx  H^3 (\eta k)^{3}\rightarrow 0$ as $\eta k\rightarrow 0$, at the end of inflation which is the signature of Bell inequality violation in our cosmological setup. Hence,
when reheating occurs after inflation a classical measure can be written down or more precisely a classical probability distribution function of fluctuation as the following:
$\rho[\phi(x)]=\mu[\phi(x)]=|\Psi[\phi(x)]|^2$. Here $|\Psi[\phi(x)]|^2$ or equivalently $\mu[\phi(x)]$ symbolizes the classical probability distribution,
which is the state of the universe at the spatial hyper surface where the process of reheating occurs.
Here all the fluctuations can be treated as distribution of classical random variables. But from the computed classical probability 
distribution function $\rho[\phi(x)]$ one cannot say anything on the exact measurement procedure on a quantum state. This is really a disappointing fact from the theoretical point of view but that is not the end of the proposal. Here it is important to mention that one can treat a measurement as a specific unitary evolution of the combined effect of physical
system as well as measuring apparatus. Such physical states can be visualized as
classical object in the context of cosmology and in a direct sense not very useful. For more details see refs.~\cite{juan:2015ja,Choudhury:2016cso}. Here we need to follow few steps to overcome the difficulties~\cite{Choudhury:2016cso}:
\begin{itemize}
\item Choose a toy model of universe that will make the
job easy.
\item Here we are not claiming that this is the unique
model of universe using which one can design the
setup.
\item Since we don’t have any direct observational
evidence we also can’t claim that this toy model is
our known universe. But may be in future this will
be tested.
\item In this toy model of universe we test the validity of
Bell’s inequality with primordial fluctuations.
\item In this computation, massive particles with spin "$s$"
($s=0,2$ and $>2$ allowed) and additionally "isospin"
quantum number plays important role.
\item We provide an example for Stringy axion which has "isospin".
\item Time dependent mass profile with dependence on
"isospin" makes the job easy. Axions have such
profile.
\item Such time dependence in mass profile of the
massive particles (ex. axion) produces classical
perturbations on the inflaton. As a result hot spots
produced in CMB by curvature fluctuations and all
such massive particles are visible today.
\end{itemize}
\section{Role of massive new particles}
\begin{figure*}[htb]
\centering
\subfigure[For $m\approx H$.]{
    \includegraphics[width=4.1cm,height=5cm] {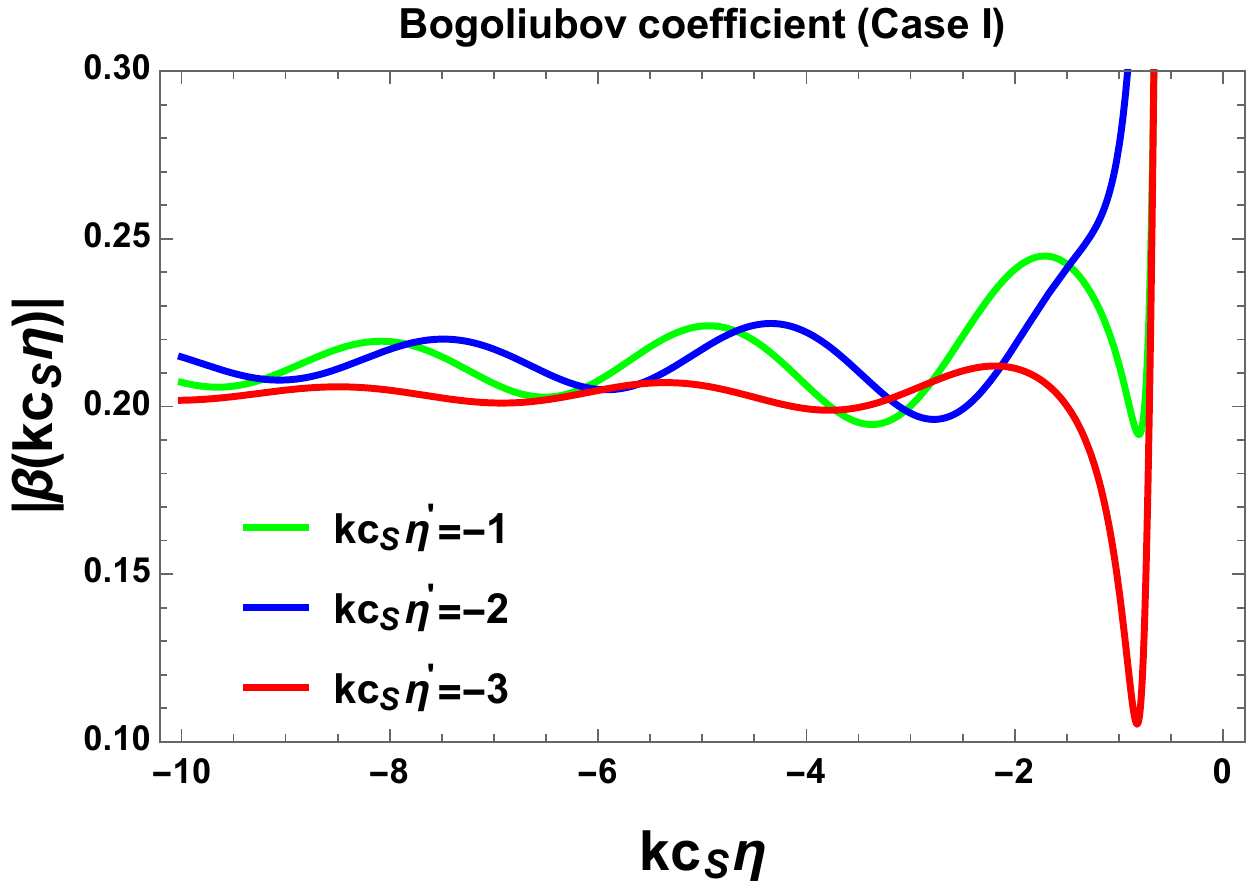}
    \label{fig1xcc1}
}
\subfigure[For $m\approx H$.]{
    \includegraphics[width=4.1cm,height=5cm] {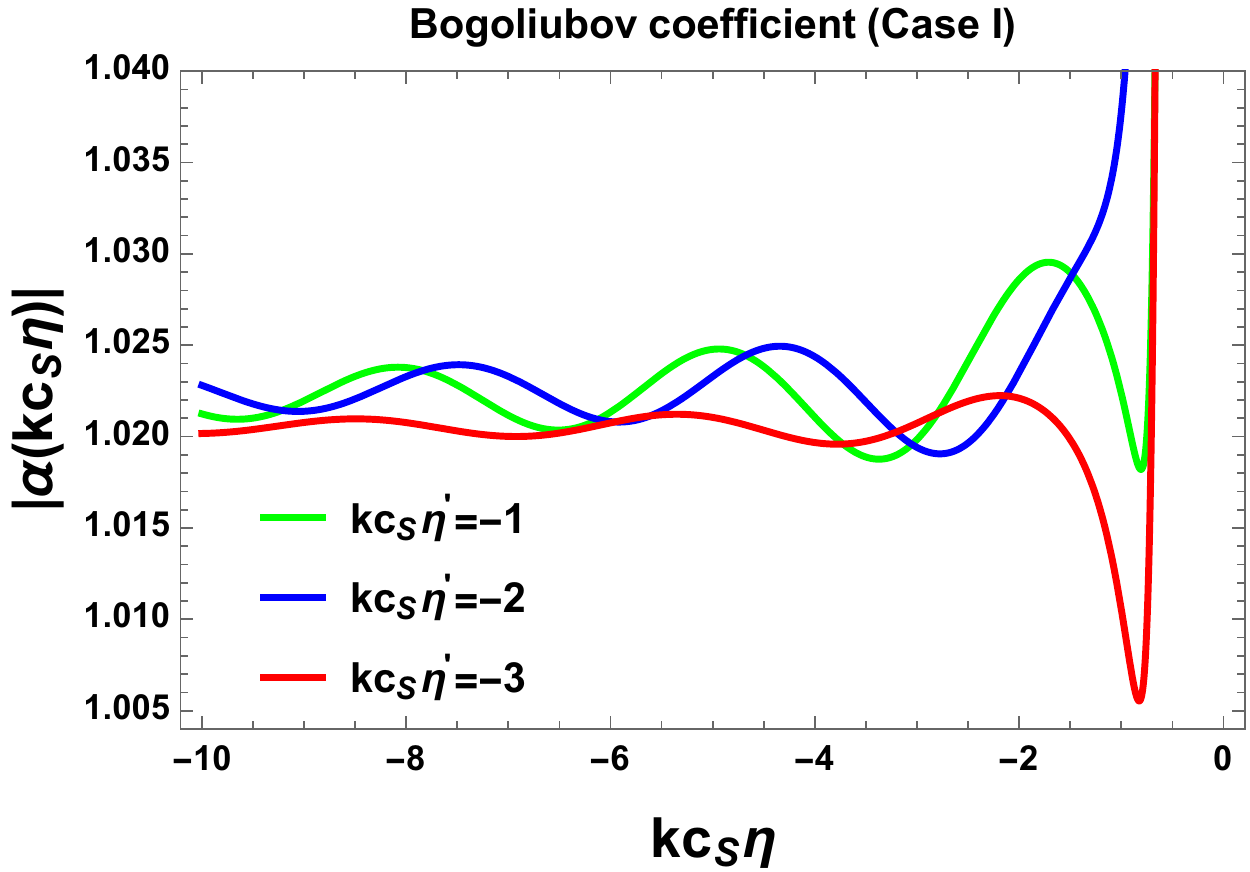}
    \label{fig2xcc2}
}
\subfigure[For $m>> H$.]{
    \includegraphics[width=4.1cm,height=5cm] {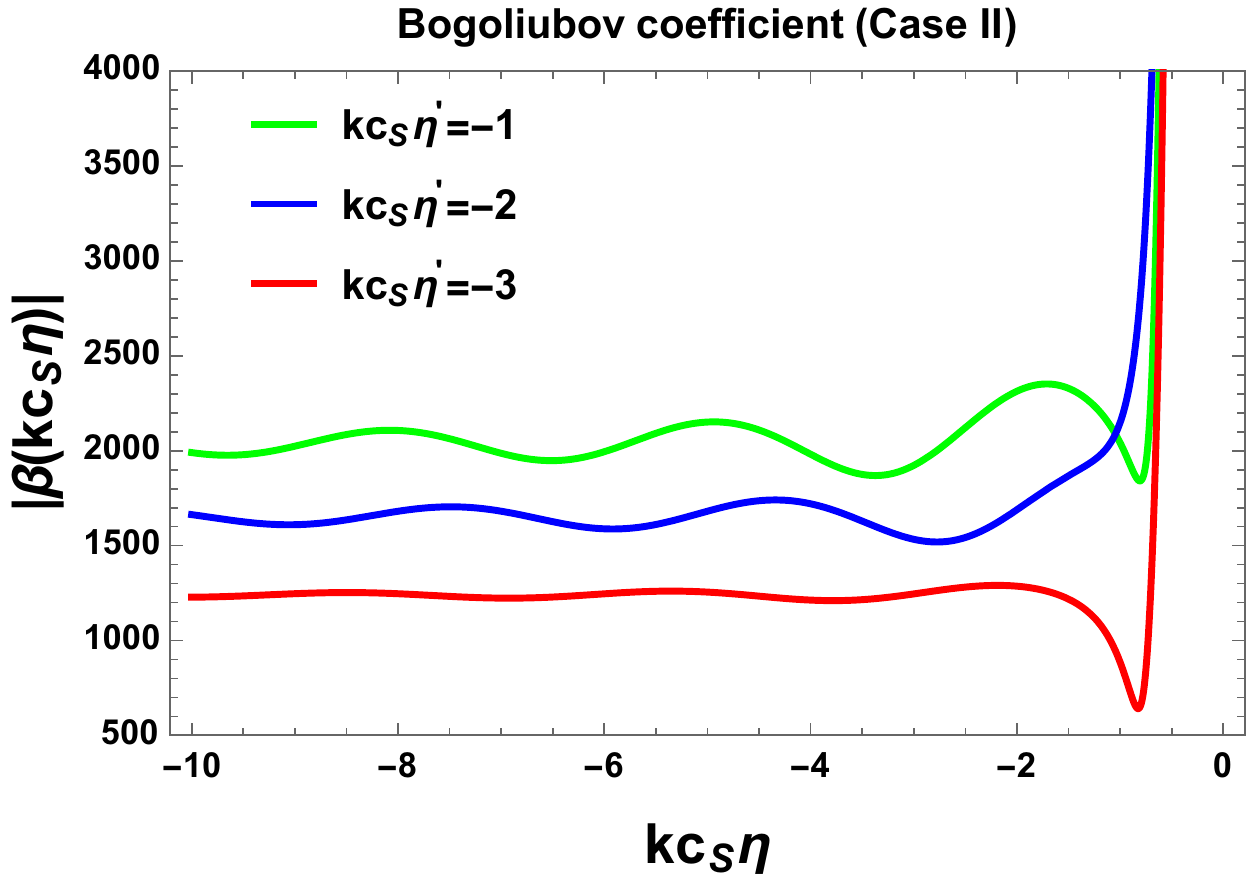}
    \label{fig3xcc3}
}
\subfigure[For $m>> H$.]{
    \includegraphics[width=4.1cm,height=5cm] {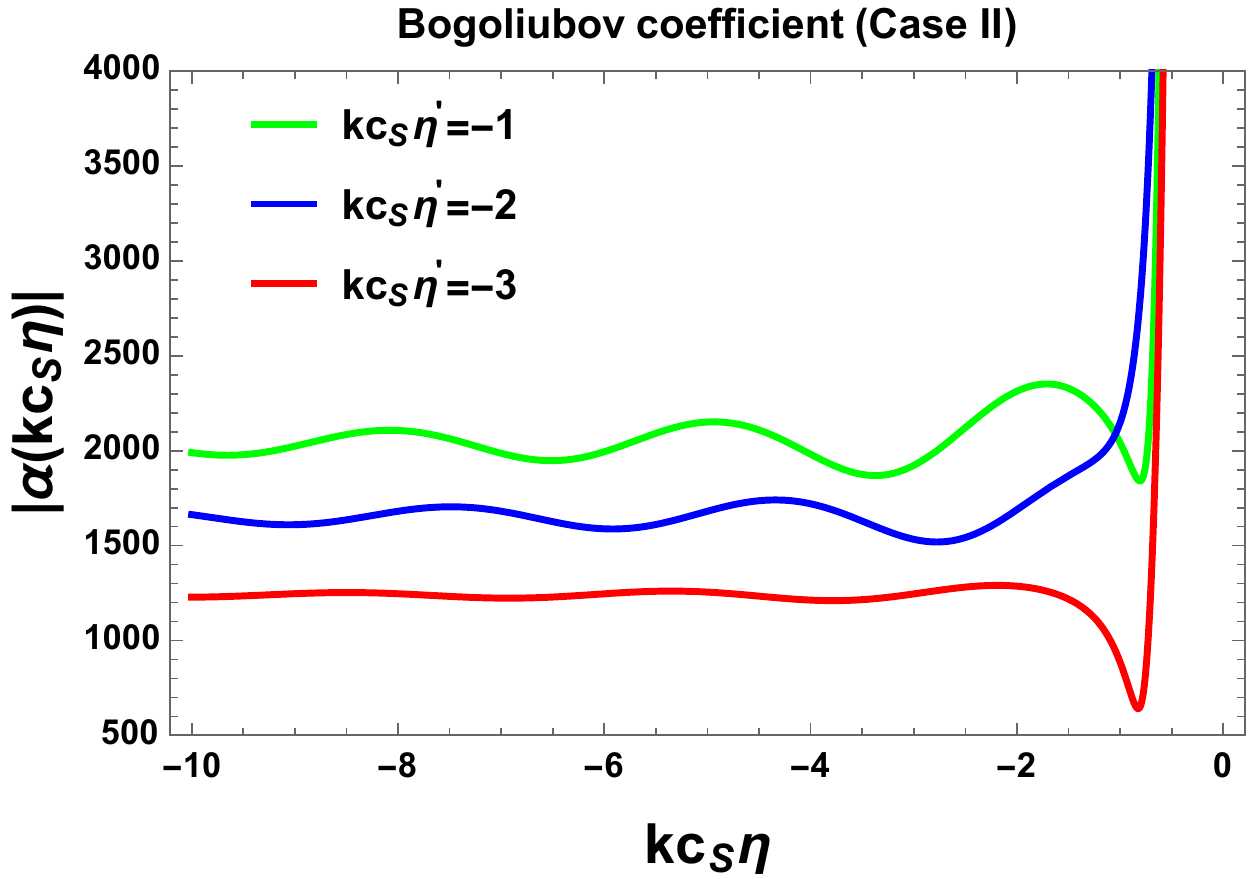}
    \label{fig4xcc4}
}
\caption[Optional caption for list of figures]{Particle creation for mass profile A for two cases- $m\approx H$ and $m>>H$.} 
\label{bogg1cc}
\end{figure*}

In this section we discuss the explicit role of massive particles in the context of Bell's inequality violation in cosmology. Here it is important to note that, all the massive particles have the few characteristics to violate Bell's inequality- A. Particle pair creation (Bell pair) is required
to produce hot and cold spots in CMB, B. Particle mass should have time
dependent profile. Growing mass profile is
usually preferred. But some times other
behavior is also important depending on
the specific mathematical structure of the
mass profile, C. Heavy fields become heavy at early and late
times, but in an intermediate time scale
where inflation occurs it behaves like a light
field, D. Each massive particle pair is created in “isospin”
singlet manner. This is required for measurement
in the detectors.
\begin{figure*}[htb]
\centering
\subfigure[Mass profile for A with $\gamma=1$ and $\delta=1$.]{
    \includegraphics[width=5.2cm,height=5cm] {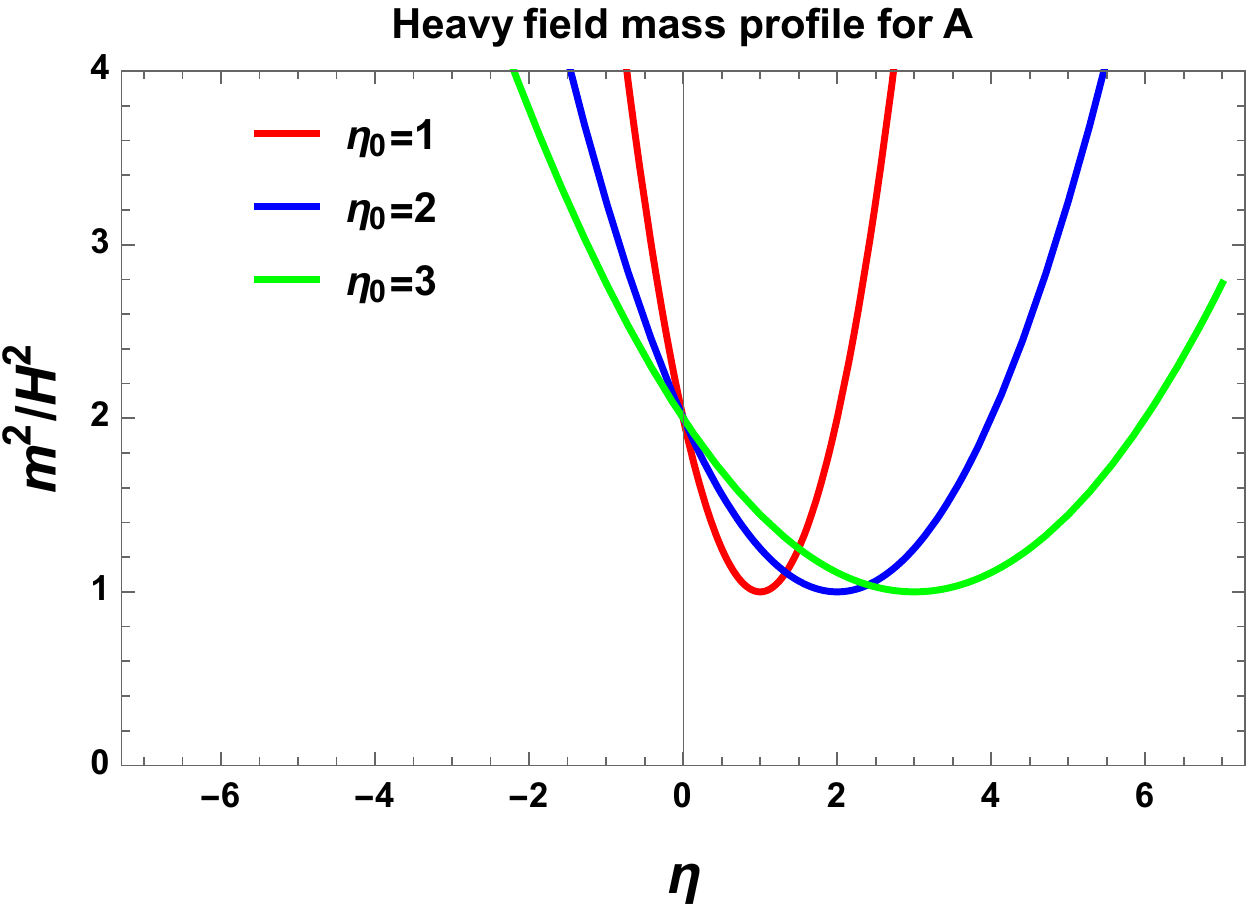}
    \label{fig1aa}
}
\subfigure[Mass profile for B with $\frac{m^{2}_{0}}{2H^2}=1$.]{
    \includegraphics[width=5.2cm,height=5cm] {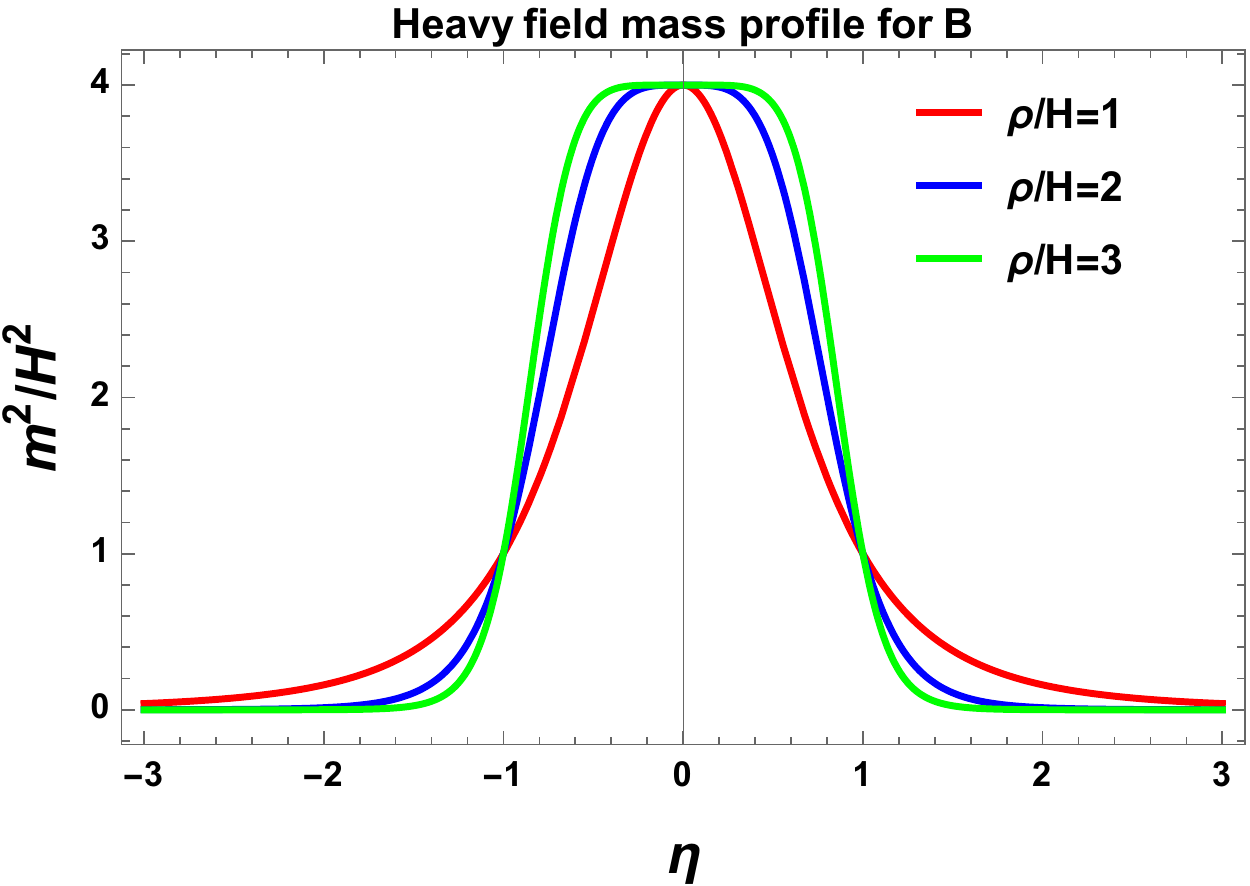}
    \label{fig2bb}
}
\subfigure[Mass profile for C with $\frac{m^{2}_{0}}{H^2}=1$.]{
    \includegraphics[width=5.2cm,height=5cm] {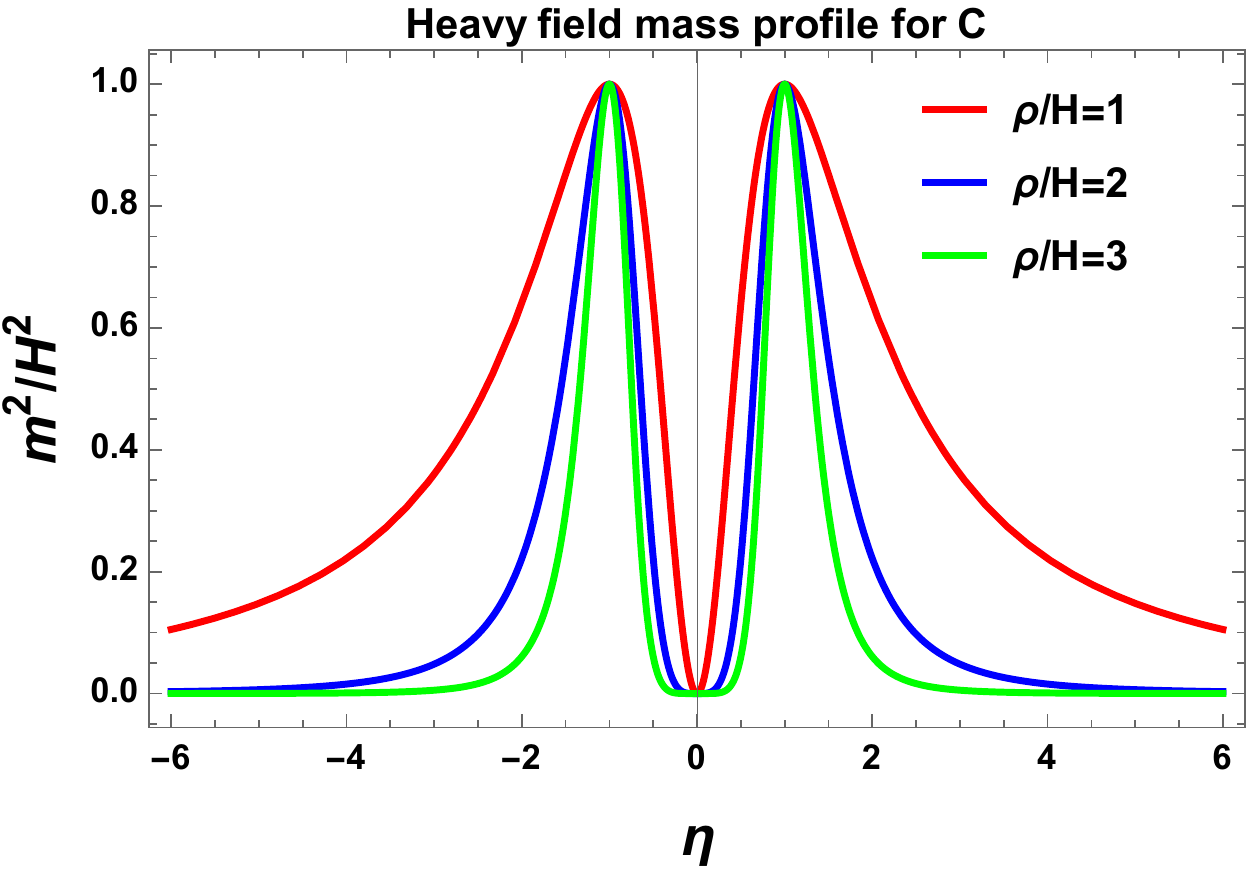}
    \label{fig2cc}
}
\caption[Optional caption for list of figures]{Toy model mass profiles for massive particles.} 
\label{faaa}
\end{figure*}
Time dependence of the inflation give rise to a time dependent mass $m(\eta)$. In the present context we consider three mass profiles as a toy model- A. $m^2/H^2= \gamma\left(\frac{\eta}{\eta_0} - 1\right)^2 + \delta$, B. $ 2m^2/m^2_0= \left[1
 -\tanh\left(\frac{\rho}{H}\ln(-H\eta)\right)\right]$ and C. $ 
  m^2/m^2_0=~{\rm sech}^2\left(\frac{\rho}{H}\ln(-H\eta)\right) $. See fig.~(\ref{faaa}), where we have shown the behaviour of all the toy models of mass profile for massive particles with conformal time scale. The equation of motion for the massive field is~\cite{Choudhury:2016cso}:
\bea
h''_k + \left\{c^2_{S}k^2 + \left(\frac{m^2}{H^2} - \left[\nu^2-\frac{1}{4}\right]
\right) \frac{1}{\eta^2} \right\} h_k &=& 0.
\eea
whereas, in the quasi de Sitter case the parameter $\nu$ can be written as, $\nu= \frac{3}{2}+\epsilon+\frac{\eta}{2}+\frac{s}{2},$
where $\eta$ and $s$ are defined as,
$\eta = \frac{\dot{\epsilon}}{H\epsilon},$
$s=\frac{\dot{c_{S}}}{Hc_{S}}$.
 Here WKB approximation is approximately valid for the mode function $h_k$ solution.
If we solve the equation of motion for the heavy fields, then we can have the solution for the
fluctuations, where it is assumed that the variation in time in heavy field mass parameter is very slow. For a case which is dependent on time, 
it is possible depending on the mathematical structure of the heavy field mass parameter $m(\eta)$.
In the standard WKB approximation the total solution is:
\bea\label{df3}
h_k (\eta)&=& \left[D_{1}u_{k}(\eta) + D_{2} \bar{u}_{k}(\eta)\right].\eea
Here $u_{k}(\eta)$ and $\bar{u}_{k}(\eta)$ are defined as:
\bea\label{solpzxzx}
 \displaystyle\small u_{k}(\eta) &=&
                    \displaystyle   \frac{1}{\sqrt{2p(\eta)}}
\exp\left[i\int^{\eta} d\eta^{\prime} p(\eta^{'})\right],~~~~
 \displaystyle\small \bar{u}_{k}(\eta) =
                    \displaystyle   \frac{1}{\sqrt{2p(\eta)}}
\exp\left[-i\int^{\eta} d\eta^{\prime} p(\eta^{'})\right]
\eea
where total solution for $h_k$ is given in terms of 
two linearly independent solutions. Here 
the new conformal time dependent factor $p(\eta)$ is defined as, $p(\eta) =\sqrt{\left\{c^{2}_{S}k^2 + \left(\frac{m^2}{H^2} - \left[\nu^2-\frac{1}{4}\right]
\right) \frac{1}{\eta^2} \right\}}$, which we have used in our calculation where $D_{1}$ and $D_{2}$ are two arbitrary integration constants and these constants depend on the
initial condition during WKB approximation at early and late time scales. In our context these
constants $D_{1}$ and $D_{2}$ can be identified with the 
Bogoliubov coefficient in momentum space,
\bea D_{1} &=& \beta(k)=\frac{{\cal R}}{{\cal T}}=\int^{0}_{-\infty}d\eta~\frac{\left(p^{'}(\eta)\right)^2}{
                    4p^3(\eta)}\exp\left[2i\int^{\eta}_{-\infty}
d\eta^{'}p(\eta^{'})\right],\\
     D_{2} &=& \alpha(k)=\frac{1}{{\cal T}}=\sqrt{1+|\beta(k)|^2}~e^{i\phi},\eea
     where ${\cal R}$ and ${\cal T}$ represent the reflection and transmission coefficient respectively. Further one can also compute the number of produced particle pairs and its energy density using following equations.:
     \bea {\cal N}_{pair}=\frac{1}{(2\pi a)^3}\int d^3{\bf k}|\beta(k)|^2,~~~~~~~\rho_{pair}=\frac{1}{(2\pi a)^3a}\int d^3{\bf k}p(\eta)|\beta(k)|^2.\eea
In the present context, to describe a very small 
fraction of particle creation after inflation, 
we have to start with a Bogoliubov coefficient $\beta$ 
in FLRW space time, which gives the amount of mixing between the WKB approximated solutions which is of two type.
Noting that, in the \textcolor{blue}{sub Hubble region} ($|kc_{S}\eta|>>1$) Bogoliubov coefficient $\beta$
is small and hence the representative probability
distribution $P(x)$ for the relative comoving distance $x$ between the two pairs peaks at the comoving length scale which is,
$x \sim |\eta_{\rm pair}|$. In the present setup, all the pair is created if typical comoving distance $x$ 
is of the order of the time $\eta_{\rm pair}$. See fig.~(\ref{bogg1cc}), where we have explicitly shown the particle creation process for mass profile A for two cases-$m\approx H$ and $m>>H$.
\section{Important note and results in cosmological perturbation theory}
Effect of the massive particles on the scalar curvature fluctuations can be described, if we take the
 second order action derived as~\cite{Choudhury:2016cso}:
\begin{equation}\label{eq1jh}
S = {\frac{1}{2}} \int d\eta d^3x \frac{2 \epsilon M^2_{p}}{c^2_{S}H^2} \left[\frac{(\partial_\eta \zeta)^2 
- c^2_{S}(\partial_i \zeta)^2}{\eta^2}-\frac{m^2_{inf}}{H^2\eta^2}\right]- \int \frac{d\eta}{c_{S}H} m(\eta) \partial_\eta \zeta(\eta,{\bf x}=0)
\end{equation}
Here the curvature perturbation $\zeta$ is related to scalar field fluctuation as, $\zeta=-H\delta\rho/\rho$ or more precisely in Fourier space $\zeta_{\bf k}=-H\phi_{\bf k}/\dot{\bar{\phi}}_0M_p$.
The last term contains the effect of massive particle as the time dependent mass parameter appears here. Additionally it symbolizes the 
 interaction term where inflaton field is interacting with the heavy fields. 
  This is a complex model of inflation
  as this model contains both inflaton and heavy field with a time dependent coupling $m(\eta)$. But in the most 
 simpler models of inflation, one can also neglect the mass contribution, 
 because $m_{inf}<<H$ approximation is valid in those cases.
We can compare our present setup with effective time varying mass 
parameter with the axions with time varying decay constant in the case
of canonical interactions which will be shown in the next section. 
 
 Using this action one can further compute the one point function as:
 \bea
 \displaystyle \langle\zeta({\bf x},\eta=0)\rangle_{|kc_{S}\eta|\rightarrow -\infty} &\approx& 
 -\frac{2H}{M^2_{p}\epsilon\pi}
 \left[|C_{2}|^2 O_{1}
 -|C_{1}|^2 O_{2}-i\left(C^{*}_{1}C_{2}e^{i\pi\left(\Lambda+\frac{1}{2}\right)}
 +C_{1}C^{*}_{2}e^{-i\pi\left(\Lambda+\frac{1}{2}\right)}\right)O_{3}\right],\\
\displaystyle \langle\zeta({\bf x},\eta=\xi\rightarrow 0)\rangle_{|kc_{S}\eta|\rightarrow 0} &=& 
 \frac{H\sqrt{\xi}c_{S}}{2M^2_{p}\epsilon\pi^2}\left(C^{*}_{1}C_{2}
 +C_{1}C^{*}_{2}-|C_{1}|^2-|C_{2}|^2\right)
 O^{\xi\theta}_{4},\\
 \langle\zeta({\bf x},\eta=\xi\rightarrow 0)\rangle_{|kc_{S}\eta|\approx 1} &=& 
 \frac{H\sqrt{\xi}c_{S}}{2M^2_{p}\epsilon\pi^2}\left(C^{*}_{1}C_{2}
 +C_{1}C^{*}_{2}-|C_{1}|^2-|C_{2}|^2\right)
 O^{\xi\theta}_{5}
 \eea
 where $O_{1}, O_{2}, O_{3}, O^{\xi\theta}_{4}, O^{\xi\theta}_{5}$ are given by:
\bea
 O_{1,2}&=& \mp\frac{i}{2\pi^2c^2_{S}}m(\eta=-|{\bf x}|/c_{S}),~~~~~ 
  O_{3}=\frac{1}{2\pi^2c^2_{S}}m(\eta=-|{\bf x}|/c_{S}),\nonumber \\
  O^{\xi\theta}_{4}&=& \frac{1}{2\pi^2}\int^{\infty}_{0} dk~k^2~e^{ikx}\int\limits_{-\infty}^{\xi} {d\eta}
 ~\frac{{c}_{S}\left(\Lambda-\frac{1}{2}\right)m(\eta)}{a(\eta)\epsilon\sqrt{-\eta}} 
 \left[\left(-\frac{kc_{S}\eta}{2}\right)^{-\Lambda}\left(-\frac{kc_{S}\xi}{2}\right)^{-\Lambda}
 +\left(\frac{kc_{S}\eta}{2}\right)^{-\Lambda}\left(\frac{kc_{S}\xi}{2}\right)^{-\Lambda}\right],~~~~~~~~~~~~\nonumber \\
  O^{\xi\theta}_{5}&=& \frac{1}{2\pi^2}\int^{\infty}_{0} dk~e^{ikx}~k^{2-\Lambda}\int\limits_{-\infty}^{\xi} {d\eta}
 ~\frac{{c}_{S}\left(\Lambda-\frac{1}{2}\right)\left(\frac{1}{2}\right)^{-\Lambda}m(\eta)}{a(\eta)\epsilon\sqrt{-\eta}}
 \left[\left(-\frac{c_{S}\xi}{2}\right)^{-\Lambda}
 +\left(-1\right)^{-\Lambda}\left(\frac{c_{S}\xi}{2}\right)^{-\Lambda}\right].~~~~~~~~~~~~\eea
Here $\xi$, is the cut-off in conformal time coordinates and acts as regulator in the theory and the parameter $\Lambda=\sqrt{\nu^2-m^2/H^2}$. Here it is important to mention that due to the presence of time dependent mass parameter $\langle\zeta({\bf x},\eta=0)\rangle$ is non zero in all the cases and precisely showing the signature of Bell's inequality violation in primordial cosmology.

Similarly in the position space the two point function is given by:
 \bea\label{dfinv4bnbncvcvcvcv}
\displaystyle \langle\zeta({\bf x},\eta)\zeta({\bf y},\eta)\rangle_{|kc_{S}\eta|\rightarrow -\infty}&\approx& 
\frac{1}{4\pi^4}\frac{H^2}{2\epsilon M^2_{p}}\frac{\eta^2\tilde{c}^2_{S}}{c_{S}}
\left[\left(|C_{2}|^2+|C_{1}|^2\right)J_{1}+\left(C^{*}_{1}C_{2}
e^{i\pi\left(\Lambda+\frac{1}{2}\right)}J_{2}
+C_{1}C^{*}_{2}e^{-i\pi\left(\Lambda+\frac{1}{2}\right)}J_{3}\right)\right],~~~~~~~~\\
\langle\zeta({\bf x},\eta)\zeta({\bf y},\eta)\rangle_{|kc_{S}\eta|\rightarrow 0}&\approx& 
\frac{H^2}{2\epsilon M^2_{p}}\frac{(-\eta c_{S})^{3-2\Lambda}}{2^{2(2-\Lambda)}\pi^4}\frac{\tilde{c}^2_{S}}{c^3_{S}}\left|\frac{\Gamma(\Lambda)}{\Gamma
\left(\frac{3}{2}\right)}\right|^2
\left[(|C_{2}|^2+|C_{1}|^2)-\left(C^{*}_{1}C_{2}
+C_{1}C^{*}_{2}\right)\right]K_{I},\\
\langle\zeta({\bf x},\eta=0)\zeta({\bf y},\eta=0)\rangle_{|kc_{S}\eta|\approx 1}&\approx& 
\frac{H^2}{2\epsilon M^2_{p}}\frac{1}{2^{2(2-\Lambda)}\pi^4}\frac{\tilde{c}^2_{S}}{c^3_{S}}\left|\frac{\Gamma(\Lambda)}{\Gamma
\left(\frac{3}{2}\right)}\right|^2
\left[(|C_{2}|^2+|C_{1}|^2)-\left(C^{*}_{1}C_{2}
+C_{1}C^{*}_{2}\right)\right]Z_{I}
\eea
where $J_{1}, J_{2}$, $J_{3}$, $K_I$ and $Z_I$ are defined as:\bea
J_{1}&=&-\frac{4\pi}{|{\bf x}-{\bf y}|^2},~~
J_{2,3}=-\frac{4\pi}{(|{\bf x}-{\bf y}|\pm 2c_{S}\eta)^2},~~K_{I}=4\pi\left(\frac{i}{|{\bf x}-{\bf y}|}\right)^{3-2\Lambda}\Gamma(3-2\Lambda),~~
Z_{I}=4\pi\left\{\ln\left(\frac{L_{IR}}{|{\bf x}-{\bf y}|}\right)-\gamma_{E}\right\}.~~~~~~~\eea
Here $1/L_{IR}$ is the infrared cut-off in the momentum scale which is introduced to regularize the momentum integral.

\section{Analogy with axion fluctuations from String Theory}
\begin{figure*}[htb]
\centering
\subfigure[Various parts of potential.]{
    \includegraphics[width=5.2cm,height=4cm] {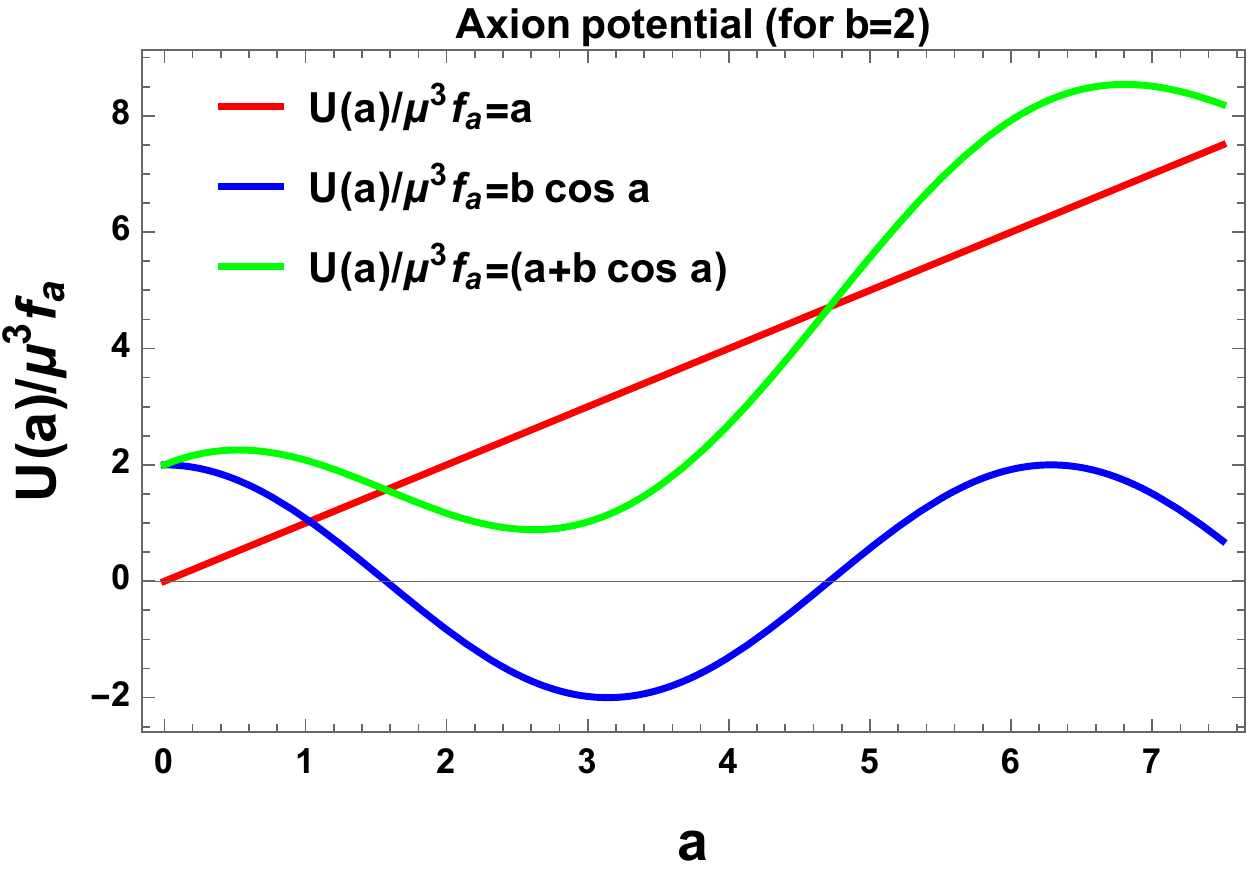}
    \label{fig1qq}
}
\subfigure[Total potential.]{
    \includegraphics[width=5.2cm,height=4cm] {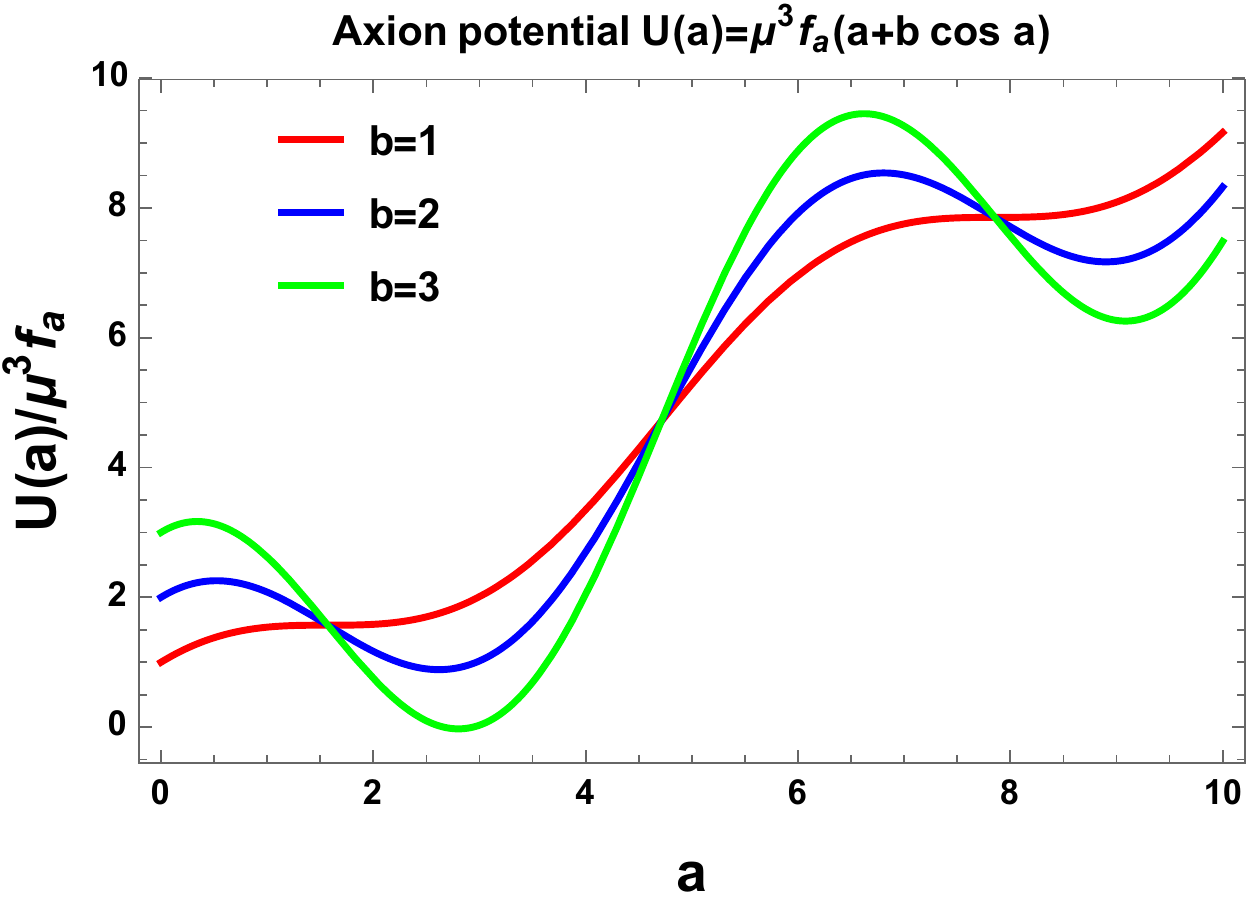}
    \label{fig2qq}
}
\subfigure[Axion decay constant profile.]{
    \includegraphics[width=5.2cm,height=4cm] {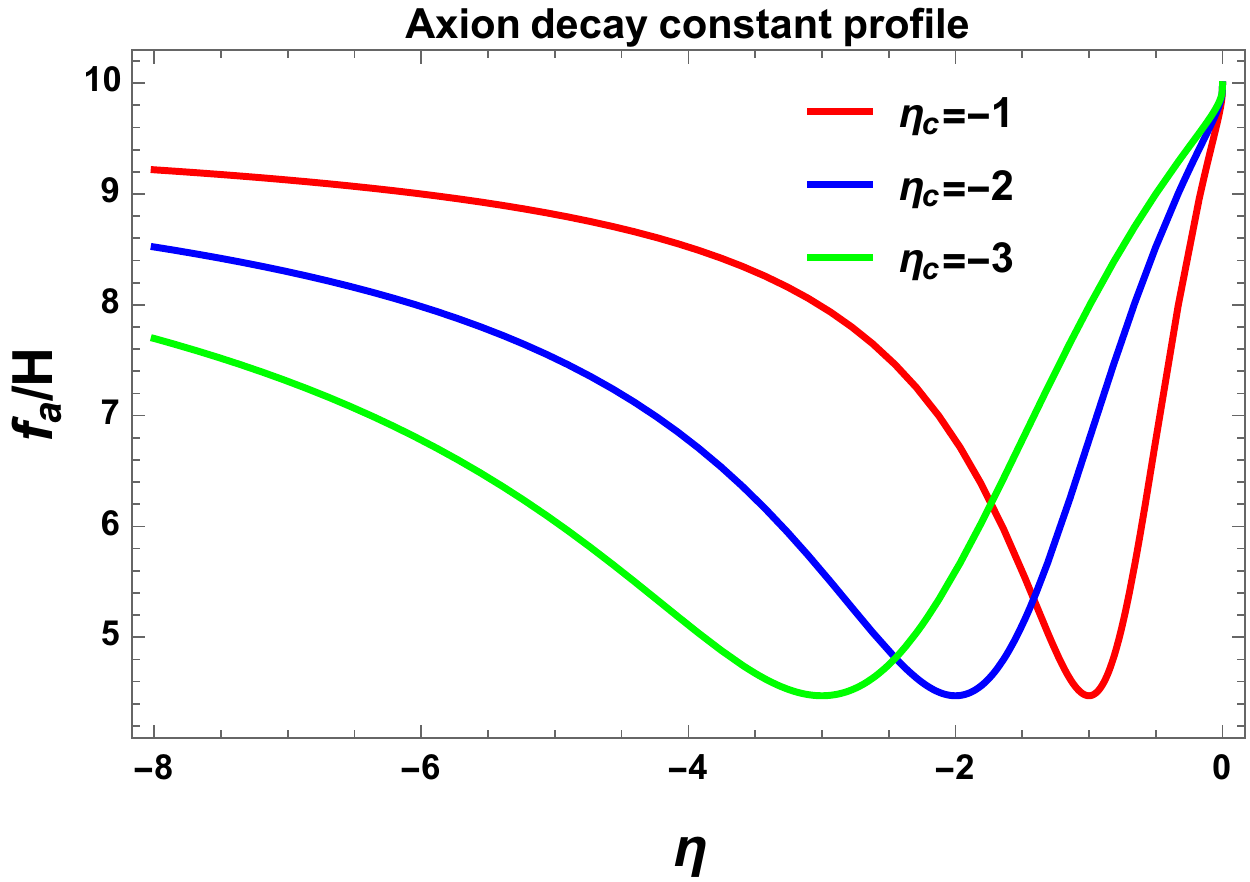}
    \label{fig3qq}
}
\caption[Optional caption for list of figures]{Behaviour of the axion effective potential and time dependent decay constant.} 
\label{vbb}
\end{figure*}
Let us consider the canonically normalized string theory originated axion action which is represented as:
\bea S_{axion}&=& \int d\eta~d^{3}x \left[\frac{f^2_{a}(\eta)}{2H^2}\frac{\left[(\partial_{\eta}a)^2-(\partial_{i}a)^2\right]}{\eta^2} -\frac{U(a)}{H^4 \eta^4}\right],\eea
where the dimensionless axion field $a=\phi/f_a$ with time dependent decay constant $f_a$ and the corresponding potential can be given by:
\bea\label{pot} U(a)&=&\mu^3af_{a}+\Lambda^4_{C}\cos a=\mu^3f_{a}\left[a+b\cos a\right].\eea
where new parameter $b$ can be represented as,
$b= \frac{\Lambda^4_{C}}{\mu^3 f_{a}}$. See fig~(\ref{vbb}), where we have shown the various parts of axion potential. Here in Eq~(\ref{pot}) the linear part of the axion potential 
has been derived in the context of string theory in ref.~\cite{Choudhury:2016cso} whereas the cosine part of the axion potential has 
its origin in non-perturbative aspects in string theory ref.~\cite{Choudhury:2016cso}.
Here, only mass contribution for axion field will contribute to the fluctuations whereas, other part can be considered as back-reaction effect which we can neglect since it is very small to be considered further.
Thus one can represent the axion potential as,
$U(a)\approx \frac{1}{2}m^{2}_{axion}(a-a_{0})^2$.
Additionally, it is important to mention that the scale $\Lambda_{C}$ is given by,~$\Lambda_{C}=\sqrt{m_{SUSY}M_{p}}~e^{-cS_{inst}}$,
       where $S_{inst}$ is the action of the instanton giving rise to effective 
       potential and, $c$ is constant factor of the order of unity,
       $m_{SUSY}$ is the supersymmetry breaking mass scale and $M_p$
       represents reduced Planck mass, which is defined as,
       $M_p=L^3/\sqrt{\alpha^{'}}g_{s}$.
In this context we assume that the axion decay constant $f_{a}$ is dependent on inflaton field and initially
this axion decay constant becomes large compared to the Hubble scale during inflation and during some time
interval, it becomes smaller compared to the Hubble scale, a few
e-foldings were created after the massive new particles, then it becomes large again. Due to the increase in
axion decay constant $f_{a}$, one can suppress quantum fluctuations effect at shorter distance scale. Hence
we get an admissible value for the magnitude of the dimensionless axion field $a$ from axion
at each created particle's location. As $f_{a}$ plays the same role as of the mass parameter $m$,
thus all the earlier computation for four cases of the choice of mass parameters are valid here. Here
the equation for the conformal time scale dependent axion decay constant is
$f_a= \sqrt{100-\frac{80}{1+\left(\ln\frac{\eta}{\eta_{c}}\right)^2}}~H$. In table~(\ref{bozxcv}) we have
shown the analogy between stringy axion and the new massive particle.

\begin{table*}
\centering
\footnotesize\begin{tabular}{|c|c|c|
}
\hline
{\bf Characteristics}& {\bf New particle} & {\bf Axion}
\\
\hline\hline\hline
 {\bf Action}&$S_{new}={\frac{1}{2}} \int d\eta d^3x \frac{2 \epsilon M^2_{p}}{\tilde{c}^2_{S}H^2} \left[\frac{(\partial_\eta \zeta)^2 
- c^2_{S}(\partial_i \zeta)^2}{\eta^2}\right]$ &  $S_{axion}=\int d\eta~d^{3}x \left[\frac{f^2_{a}(\eta)}{2H^2}\frac{\left[(\partial_{\eta}a)^2-(\partial_{i}a)^2\right]}{\eta^2} -\frac{U(a)}{H^4 \eta^4}\right].$
\\
  & $- \int \frac{d\eta}{\tilde{c}_{S}H} m(\eta) \partial_\eta \zeta(\eta,{\bf x}=0)$ &  $- \int \frac{d\eta}{f_a H} m_{axion} \partial_\eta \bar{a}(\eta,{\bf x}=0)$
\\
\hline
{\bf Mass parameter}  &$\frac{m^2}{H^2}=\tiny
\left\{\begin{array}{ll}
                    \displaystyle \gamma\left(\frac{\eta}{\eta_0} - 1\right)^2 + \delta~~~~ &
 \mbox{\small {\bf ~Case~I}}  \\ 
 \displaystyle \frac{m^2_0}{2H^2}\left[1-\tanh\left(\rho\frac{\ln(-H\eta)}{H}\right)\right]~~~~ &
 \mbox{\small {\bf ~Case~II}}  \\ 
	\displaystyle \frac{m^2_0}{H^2}{\rm sech}^2\left(\rho\frac{\ln(-H\eta)}{H}\right) ~~~~ 
	& \mbox{\small {\bf ~Case~III}}.
          \end{array}
\right.
$  &   $\frac{m_{axion}}{f_{a}}=\tiny\left\{\begin{array}{ll}
                    \displaystyle \sqrt{-\frac{\Lambda^4_C}{f^2_a} \cos\left(\sin^{-1}\left(\frac{\mu^3 f_{a}}{\Lambda^4_{C}}\right)\right)}~~~~ &
 \mbox{\small {\bf for ~total~$U(a)$}}  \\ \\
	\displaystyle \sqrt{\frac{\Lambda^4_C}{f^2_a}(-1)^{m+1}} ~~~~ & \mbox{\small {\bf for~osc.~$U(a)$}}.
          \end{array}
\right.$\\
\hline 
{\bf Rescaled mode}  &$h_{\bf k}=-\frac{\sqrt{2\epsilon}}{H\eta\tilde{c}_{S}}M_{p}\zeta_{\bf k}$ &   $\vartheta_{\bf k}=\frac{f^2_a}{H^2\eta^2M^2_p}\bar{a}_{\bf k}$\\
\hline
{\bf Scalar mode }  &$h^{''}_{\bf k}+
\left(c^2_{S}k^2+\frac{\left(\frac{m^2}{H^2}-\delta\right)}{\eta^2}\right)h_{\bf k}=0$
&   $\partial^{2}_{\eta}\vartheta_{\bf k}+
\left(k^2-\frac{\partial^2_{\eta}\left(\frac{f^2_{a}}{H^2\eta^2}\right)}{\left(\frac{f^2_{a}}{H^2\eta^2}\right)}
+\frac{m^{2}_{axion}}{f^2_{a}H^2\eta^2}\right)\vartheta_{\bf k}=0$\\
{\bf equation} & where $\delta=\frac{z^{''}}{z}=\small\left\{\begin{array}{ll}
                    \displaystyle 2~~~~ &
 \mbox{\small {\bf for ~dS}}  \\ 
	\displaystyle \left(\nu^2-\frac{1}{4}\right) ~~~~ & \mbox{\small {\bf for~qdS}}.
          \end{array}
\right.$& where $\frac{\partial^2_{\eta}\left(\frac{f^2_{a}}{H^2\eta^2}\right)}{\left(\frac{f^2_{a}}{H^2\eta^2}\right)}
\approx \tiny
\left\{\begin{array}{ll}
                    \displaystyle \frac{6}{\eta^2}~~~~ &
 \mbox{\small {\bf for ~$\eta\sim \eta_{c}$}}  \\ 
 \displaystyle \frac{6}{\eta^2}~~~~ &
 \mbox{\small {\bf for ~early~\&~late~$\eta$}}  \\ 
	\displaystyle \frac{6+\Delta_{c}}{\eta^2} ~~~~ & \mbox{\small {\bf for~$\eta<\eta_{c}$}}.
          \end{array}
\right.$\\ 
\hline 
\end{tabular}
\caption{ Table showing the analogy between the new particle and axion in string theory.}\label{bozxcv}
\vspace{.4cm}
\end{table*}
\section{Role of isospin breaking interaction and detection}
\begin{figure*}[htb]
\centering
\subfigure[Here we set $\gamma_{\bf even}=1=\delta_{\bf even}$, $\gamma_{\bf odd}=0.5=\delta_{\bf odd}$.]{
    \includegraphics[width=5.2cm,height=4cm] {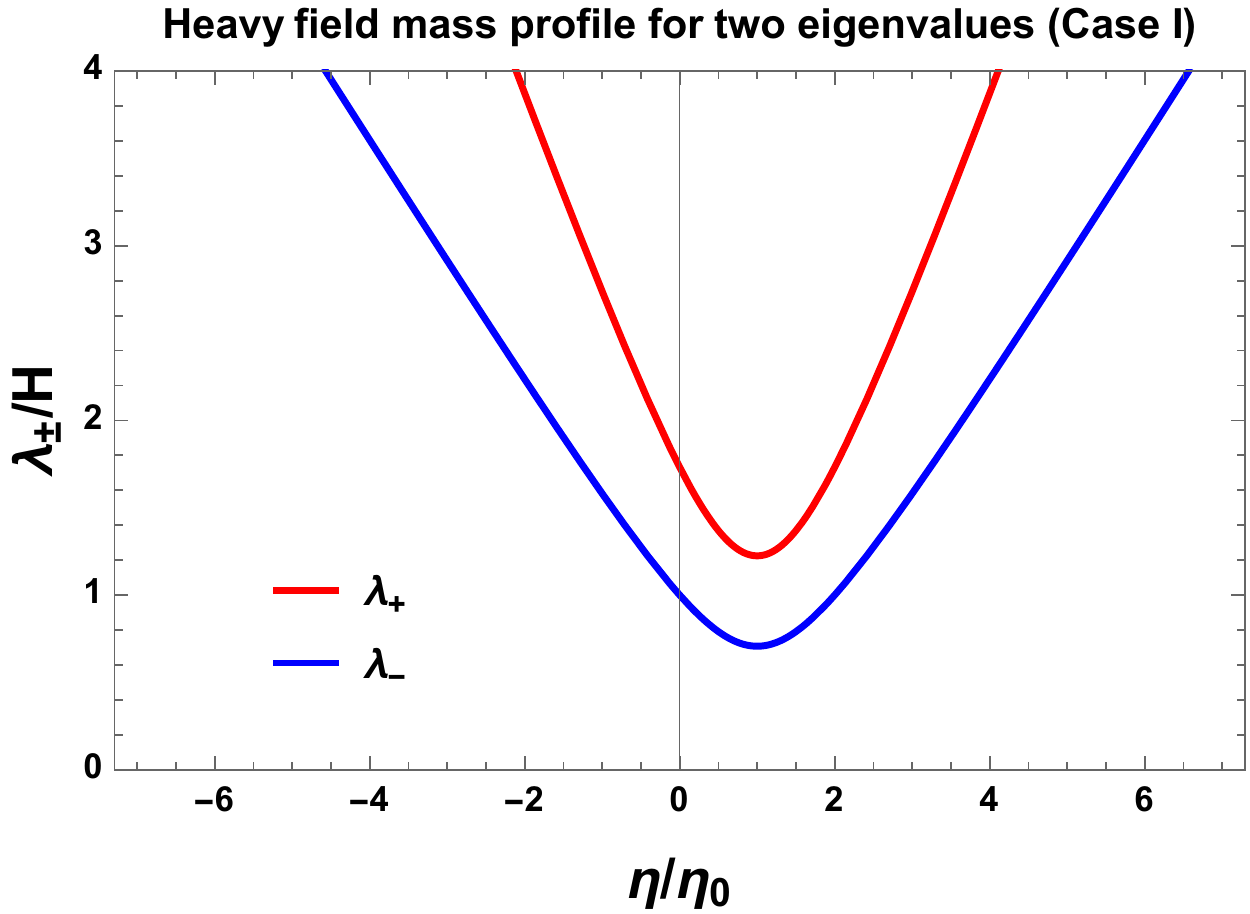}
}
\subfigure[Here we set $m_{\bf even}=3$, $m_{\bf odd}=1.5$ and $\rho/H=1$.]{
    \includegraphics[width=5.2cm,height=4cm] {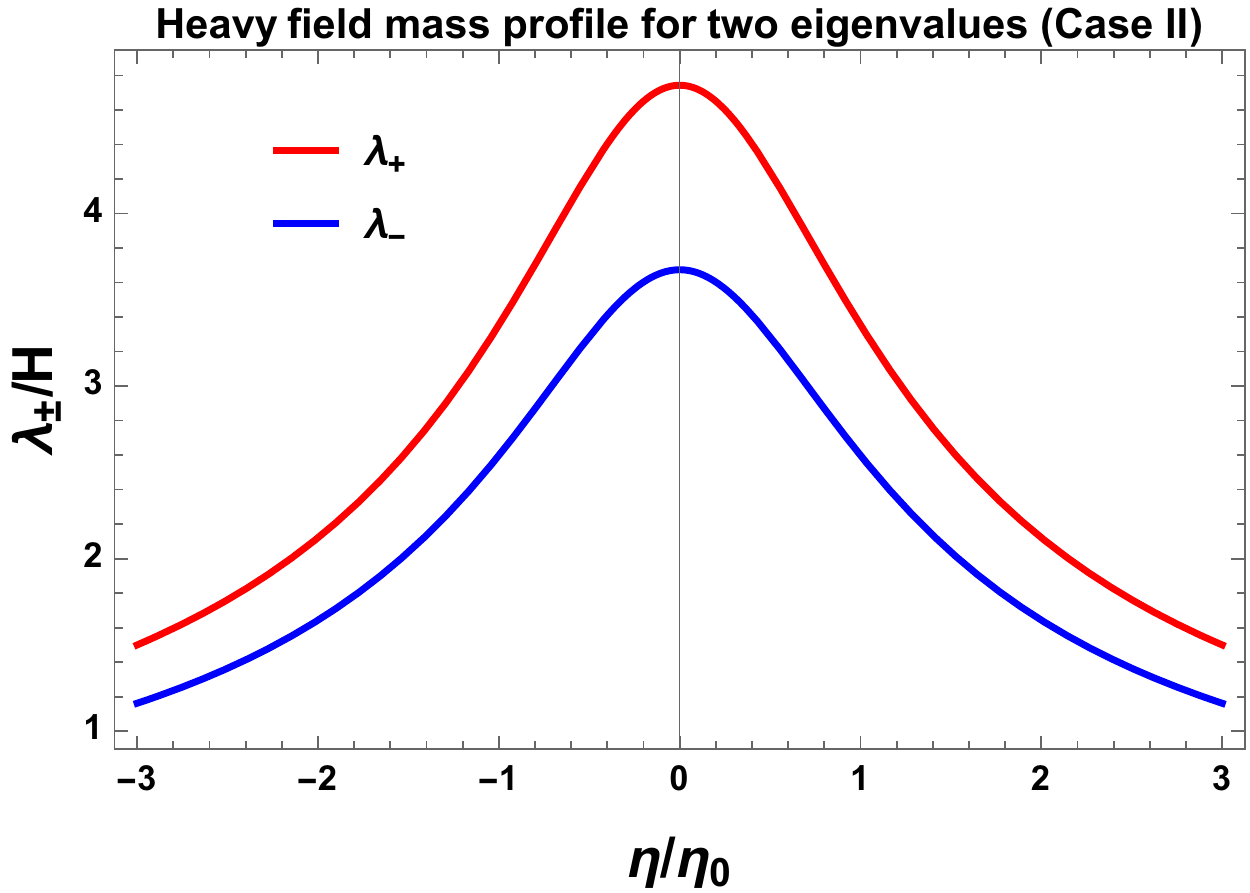}
}
\subfigure[Here we set $m_{\bf even}=2$, $m_{\bf odd}=1.5$ and $\rho/H=1$.]{
    \includegraphics[width=5.2cm,height=4cm] {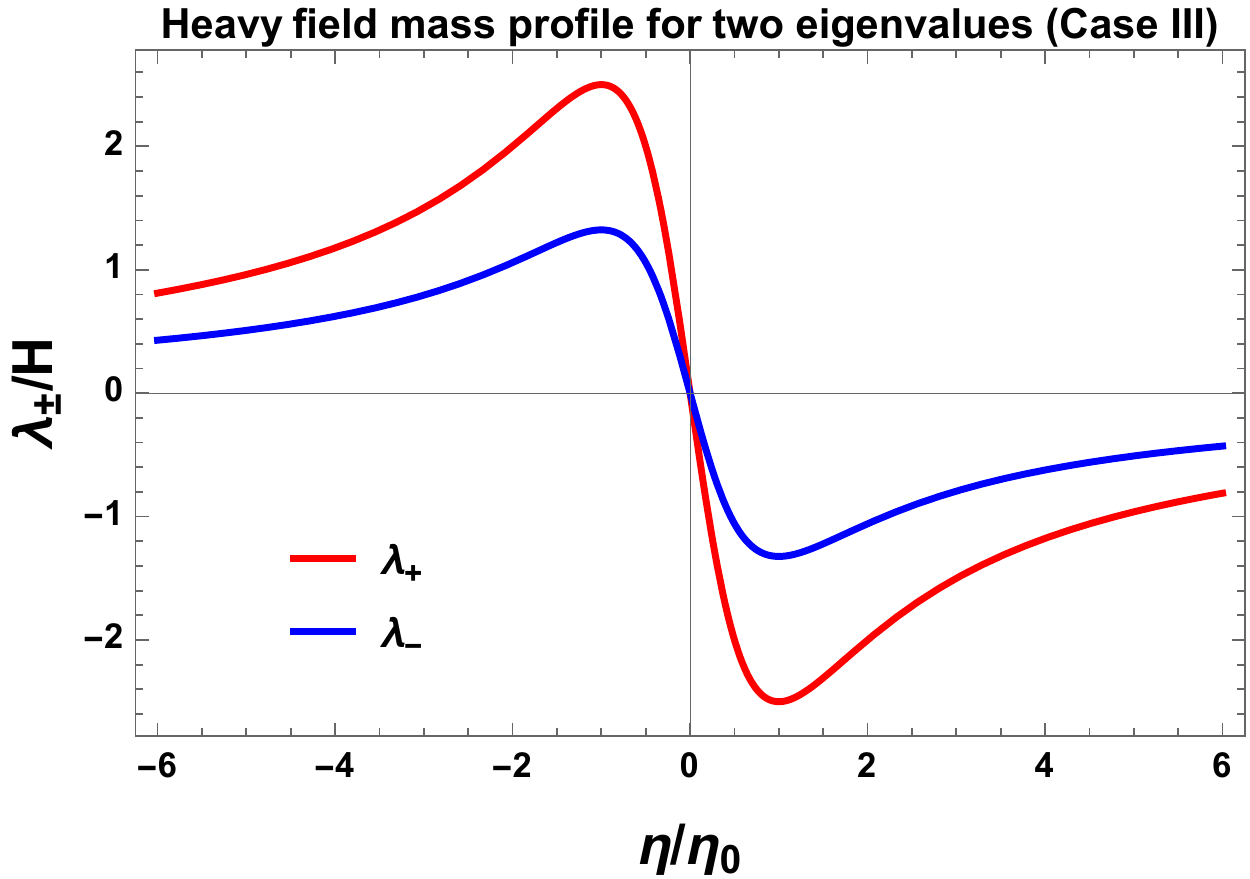}
}
\caption[Optional caption for list of figures]{Conformal time scale dependent behaviour of heavy particle mass profile for two eigenstates.} 
\label{hmhm}
\end{figure*}
We know that to construct a Bell violating experiment in the context of cosmology, isospin plays a very significant role. Therefore, let us consider following
effective Lagrangian for the inflaton and heavy scalar field interaction which is given by:
\bea\label{axhev} S
&=&\int d^{4}x\sqrt{-g}\left[\frac{1}{2}g^{\mu\nu}(\partial_{\mu}{\cal H}_{1})^{*}(\partial_{\nu}{\cal H}_{1})+\frac{1}{2}g^{\mu\nu}(\partial_{\mu}{\cal H}_{2})^{*}(\partial_{\nu}{\cal H}_{2})+\frac{1}{2}g^{\mu\nu}(\partial_{\mu}\phi)(\partial_{\nu}\phi)
\right.\nonumber\\ &&\left.~~~~~ -\left[|{\cal H}_{1}|^2+|{\cal H}_{2}|^2\right]
{\cal M}^{2}_{\bf even}(\phi)-\left[\exp(-im\theta){\cal H}^{*}_{1}{\cal H}_{2}+\exp(im\theta){\cal H}^{*}_{2}H_{1}\right]
{\cal M}^{2}_{\bf odd}(\phi)+\cdots\right].\eea
 Here $\phi$ is inflaton field and the heavy field has a isospin $SU(2)$ doublet which is given by,
${\cal H}=({\cal H}_{1},{\cal H}_{2})$. 
Here the mass-terms are defined as,
${\cal M}^{2}_{\bf even}(\phi)=\sum^{\infty}_{n=0,2,4}{\cal M}^{2}_{n}(\phi)$,
${\cal M}^{2}_{\bf odd}(\phi)=\sum^{\infty}_{n=1,3,5}{\cal M}^{2}_{n}(\phi)$. Using this, one can construct a physical mass eigen basis from the eigenvalues:
$\lambda_{\pm}(\phi)=\sqrt{{\cal M}^{2}_{\bf even}(\phi)\pm {\cal M}^{2}_{\bf odd}(\phi)}$.
To show the time dependent behaviour of the mass parameters which are constructed out of the {\bf even} and {\bf odd} contributions in the interaction picture here we chose, I. $ \lambda_{\pm}(\eta)/H= \sqrt{\gamma_{\pm}\left(\frac{\eta}{\eta_0} - 1\right)^2 +\delta_{\pm}}$, II. $   \lambda_{\pm}(\eta)/H=\frac{m_{0\pm}}{\sqrt{2}H}\sqrt{\left[1-\tanh\left(\rho\frac{\ln(-H\eta)}{H}\right)\right]}$, III.  
	$\lambda_{\pm}(\eta)/H=\frac{m_{0\pm}}{H}{\rm sech}\left(\rho\frac{\ln(-H\eta)}{H}\right)$.
	To simplify the calculations, we introduce new parameters as:
$\gamma_{\pm}= \gamma_{\bf even}\pm \gamma_{\bf odd}$, $
\delta_{\pm}=\delta_{\bf even}\pm \delta_{\bf odd}$,
$m_{0\pm}=\sqrt{{m}^2_{\bf even}\pm {m}^2_{\bf even}}$.
See fig~(\ref{hmhm}), where we have shown the conformal time dependent behaviour of three toy models of heavy particle mass profile for two eigen states. 
To make the eigen basis stable, eigen values of the mass matrix are always positive definite. At late time scales ${\cal M}^{2}_{\bf even}(\phi) \sim {\cal M}^{2}_{\bf odd}(\phi)$. Hence, the eigen value of the mass-matrix
increases. Another important requirement is, eigen values of the mass matrix should be of the order of $M_p$. The eigen value
of the mass matrix is , $ \lambda(\phi)\sim {\cal M}_{\bf even}(\phi)$, when we know that all the isospin breaking interactions are absent from the effective Lagrangian. This is also one criteria of significant importance that
can be observed with $SU(2)$ isospin singlet state during the time of heavy mass particle creation. Here the angular parameter $\theta$ and its functional dependence on the background plays extremely important role to setup the Cosmological Bell violating setup.
To make the case easy, one can assume that $\theta$ is a constant. Here, if the particle mass eigen value is $\lambda_{\pm}(\phi)$, then the antiparticle mass eigen values are given by 
$\lambda_{\pm}(\phi)$. The sign of the eigen value of the antiparticle mass eigen state may flip if the angular parameter $\theta$ is not a constant quantity but background dependent.
After the period of inflation, we can set the axion potential to be zero to avoid the problem of domain wall
formation. We have not yet observed any signatures of these heavy fields or the axion, therefore we can
consider such heavy fields or the axions corresponding to a component of dark matter and we can treat its corresponding density fluctuations as isocurvature fluctuations.

\section{Role of spin for new particles}
Let us generalize the idea where we consider the situation for a dynamical massive field with arbitrary spin ${\cal S}$. Therefore in this case, the classical time dependence of the high spin modes give rise
to time dependent mass $m_{\cal S}\left(\eta\right)$ for the spin field. Hence, for the massive field with arbitrary spin ${\cal S}$, the equation of motion is given by:
\bea
h''_k + \left\{c^2_{S}k^2 + \left(\frac{m^2_{\cal S}}{H^2} - \left[\nu^2_{\cal S}-\frac{1}{4}\right]
\right) \frac{1}{\eta^2} \right\} h_k &=& 0.
\eea
where in quasi de-Sitter case, 
$\nu_{\cal S}= \left({\cal S}-\frac{1}{2}\right)+\epsilon+\frac{\eta}{2}+\frac{s}{2}$. Now following the same procedure mentioned earlier finally we give the estimate of the mass parameter as, $ \left|\frac{m_{\cal S}}{H}\right|= \left|\sqrt{\nu^2_{\cal S}-\frac{(4-n_{\zeta})^2}{4}}\right|\geq  \sqrt{\nu^2_{\cal S}-\frac{1}{4}}$, where $n_{\zeta}$ is
scalar spectral tilt, which is constarined from cosmological observation (Planck 2015 data).

\section{Conclusion}
To summarize, in the present article we have addressed following important points:
\begin{itemize}
\item We provide a toy model for Bell's inequality
violation in cosmology.
\item Model consist of inflaton and additional massive
field with time dependent behaviour.
\item For each model, massive particle creation in
"isospin" singlet state plays a crucial role.
\item Prescribed methodology is consistent with axion
fluctuations appearing in the context of String
Theory.
\item Signature of the Bell violation is visualized from
non-zero one point function of curvature
fluctuation. Additionally we provide the result of two point function which has direct observational consequence.
\item Finally, we also provide the mass bound on the new particle in terms of the arbitrary spin ${\cal S}$.
\end{itemize}

\vspace{6pt} 

\appendix{}

\section*{Acknowledgments}
SC would like to thank Department of Theoretical Physics, Tata Institute of Fundamental
Research, Mumbai for providing me Visiting (Post-Doctoral) Research Fellowship. SC would like to thank Gautam Mandal and Guruprasad Kar for useful discussions and suggestions. SC take this opportunity to thank
sincerely to Sandip P. Trivedi and Shiraz Minwalla for their constant support
and inspiration.  
SC also thanks the organizers of VARCOSMOFUN'16 where this work was presented.
Last but not the least, we would all like to acknowledge our debt to the people of India for their generous and steady support for research in natural
sciences, especially for string theory and cosmology.

\renewcommand\bibname{References}

\end{document}